\documentclass[twocolumn,amsmath,nofootinbib,amssymb]{revtex4}
\usepackage{epsfig}
\usepackage{graphicx}% Include figure files
\usepackage{dcolumn}% Align table columns on decimal point
\usepackage{bm}% bold math
\usepackage{color}

\newcommand{\Od}{{\cal O}}
\def\thebiblio#1{
\begin{center}\bf \large References
\end{center}
\list
{[\arabic{enumi}]}{\settowidth\labelwidth{#1.}\leftmargin\labelwidth
 \advance\leftmargin\labelsep
 \usecounter{enumi}}
 \def\newblock{\hskip .11em plus .33em minus -.07em}
 \sloppy
 \sfcode`\.=1000\relax}

\begin{document}
\preprint{}
\title{%
Isotropy theorem for cosmological vector fields
}
% Force line breaks with \\

\author{J. A. R. Cembranos, C. Hallabrin,  A.\,L.\,Maroto and S. J. N\'u\~nez Jare\~no}
\address{Departamento de  F\'{\i}sica Te\'orica I, Universidad Complutense de Madrid, E-28040 Madrid, Spain}

\date{\today}% It is always \today, today,
             % but any date may be explicitly specified

\begin{abstract}
We consider homogeneous abelian vector fields in an expanding universe.
We find a mechanical analogy in which the system behaves as a particle moving in three dimensions
under the action of a central potential. In the case of  bounded and rapid evolution compared
to the rate of expansion,  we show by making use of the virial theorem
that for arbitrary potential and polarization pattern, the average energy-momentum tensor is always
diagonal and isotropic despite the intrinsic anisotropic evolution of the vector field.
For simple power-law potentials  of the form $V=
\lambda  (A^\mu  A_\mu)^n$, the average equation of state is found to be  $w=(n-1)/(n+1)$.
This implies that vector coherent oscillations could act as natural dark matter or dark energy candidates. Finally, 
we show that under very general conditions, the average energy-momentum tensor of
a rapidly evolving bounded vector field in any background geometry is always isotropic
 and  has the perfect fluid form for any locally inertial observer.
\end{abstract}

\maketitle

%\section{Introduction}

Our knowledge about the history of our universe has improved over the last years
with the advent of a large amount of new observations. There are robust
astrophysical data that support the existence of an early inflationary era; 
an additional matter component supplementing the baryonic one, known as dark matter; 
and a present era of accelerated expansion driven by dark energy.
However, the fundamental nature of these components remains
unknown. Different phenomenological and model building strategies have been proposed 
to understand them in which rapidly evolving fields could play an important role.

In particular, different scalar dynamics have been considered as the possible answers
for these open questions: standard inflation models are based on the slow-roll evolution
of a scalar field known as inflaton. The dynamics of this field is able to finish successfully
the accelerated regime and its later rapid oscillations provide a  mechanism for reheating, transferring
perturbatively or non-perturbatively the energy density of the oscillations to the matter fields.
Also the possibility of generating inflationary expansion during the period of inflaton oscillations
has been studied in \cite{Damour}. Another example of rapidly evolving scalar fields can be found
in non-thermal dark matter candidates like the axion \cite{axions}
or other massive scalar \cite{scalars} or pseudoscalar particles \cite{branons}, in which the energy density
of rapid scalar coherent oscillations scales precisely as non-relativistic matter \cite{Turner}.
Dark energy models based on the dynamics of scalar fields are commonly known as quintessence. Oscillating
evolutions have also been considered within this context \cite{Liddle}.

On the other hand, it is interesting to remark that all these possibilities could
in principle be offered by any bosonic degree of freedom and not
only by scalars. In fact, a large number of fundamental vector fields are present in the standard 
model of particles and interactions, and in the
most part of its extensions. Therefore it is natural to consider vector models which could
shed light on the above mentioned open problems in cosmology. However, there is an
important distinctive feature of vector fields as compared with scalar fields.
Even in a homogeneous configuration, vector dynamics is generally anisotropic.

Most part of observational data seem to be consistent with an early homogenous and isotropic universe.
Models supporting a large amount of anisotropy suffer severe constraints. However, despite this
fact, there are some examples in which vector fields have been shown to provide interesting models
in different cosmological scenarios.  For instance, models of inflation based on vector fields
have been extensively studied recently \cite{vectorinflation}, and a first proposal for vector
inflation which could avoid the generation of an excess of anisotropy can be found in \cite{Ford}.
Models of dark energy based on massive vector fields have been also considered in \cite{DE}.
Vector dark energy without potential terms has been proposed in \cite{VT}.
Vector models for dark matter based on hidden sector gauge bosons have been discussed in \cite{VDM}.
A possible role in the generation of metric perturbations in the so called curvaton scenario,
has also been considered for vectors in \cite{Dimopoulos}. In that work, it was shown that
an oscillating massive homogeneous vector field behaves as non-relativistic matter
with an equation of state $\omega=0$, in a completely analogous way to the scalar case.
Despite the anisotropy of the oscillations, the average energy-momentum tensor (EMT)
turned out to be isotropic in that case. This fact has been used in order to propose 
oscillating massive vector
fields as  non-thermal dark matter candidates in \cite{Nelson}.

On the other hand, the potential of anisotropic cosmologies have received important attention
in the last years, mainly motivated by the  possible existence of anomalies in the
isotropy of the Cosmic Microwave Background (CMB) and matter distributions which could be pointing to the existence
of a preferred spatial direction in the Universe \cite{anomalies}.

In this work, we show that despite its intrinsically anisotropic evolution, the average EMT associated to rapidly evolving vector fields is isotropic
under very general and natural conditions. The proof has a mechanical analogy with the virial theorem, and applies
to a general Abelian gauge vector with any general self-interaction given by a potential of the form $V(A_\mu A^\mu)$.

%\section{Oscillating vector fields in a FLRW background}\label{eom}
In order to simplify the argument, let us consider first a flat Friedmann-Lema\^itre-Robertson-Walker
(FLRW) metric given by:
\begin{eqnarray}
ds^2=dt^2-a^2(t) d\vec x^2\;.
\end{eqnarray}
The Lagrangian density for a vector field with a potential which is an arbitrary scalar
function of $A^2=A_\mu A^\mu$ is given by:
\begin{eqnarray}
{\cal L}=-\frac{1}{4}F_{ \mu\nu}F^{\mu\nu}-V (A^2)\;,
\label{L}
\end{eqnarray}
where, for an Abelian field, the field strength tensor is
\begin{equation}
F_{\mu\nu}=
%\nabla_\mu A_\nu -\nabla_\nu A_\mu =
\partial_\mu A_\nu - \partial_\nu A_\mu
%\;,
\;.
\label{F}
\end{equation}
%with $\nabla_\mu$ being the covariant derivative.
%The Euler-Lagrange equations are:
%%
%\begin{equation}
%\nabla_\mu\left[\frac{\partial{\cal L}}{\partial(\nabla_\mu A_\nu)}\right]=
%\frac{\partial{\cal L}}{\partial A_\nu}\,.
%\label{EL}
%\end{equation}
 The corresponding field equations read:
\begin{equation}
F^{\mu \nu}_{\;\;\;\; ;\nu} +2 V'(A^2) A^\mu = 0\;,
\end{equation}
where $V'(x)=dV/dx$. We also need to calculate the EMT:
\begin{eqnarray}
T^\mu _{\;\;\nu}&=&\frac{1}{4} F_{\rho \lambda} F^{\rho \lambda}  g^\mu_{\;\;\nu} - F^{\rho \mu} F_{\rho \nu}
\nonumber\\
&+&  V(A^2)g^\mu_{\;\;\nu}
-2V' (A^2) A^\mu A_\nu\;.
\end{eqnarray}
Assuming homogeneity for the vector field, i.e. $A_\mu=(A_0(t),A_i(t))$, we obtain for the
temporal component:
\begin{eqnarray}
V'(A^2)A_0=0\;, \label{temporal}
\end{eqnarray}
whereas for the spatial components, we get:
\begin{equation}
\ddot{A_i} + H \dot{A_i} - 2 V'(A^2)A_i = 0\;. \label{motion}
\end{equation}

The EMT components read:
\begin{eqnarray}
 \rho &\equiv& T^0_{\;\;0}= \frac{1}{2} \frac{\dot A_i\dot A_j \delta^{ij}}{a^2}  +V (A^2)\;\;; \label{energy}\\
p_k &\equiv& - T^k_{\;\;k } = \frac{1}{2} \frac{\dot{A_i} \dot{A_j}}{a^2} \delta^{i j} - \frac{\dot{A_k} \dot{A_k}}{a^2}  \nonumber\\
&-& V(A^2) - 2V'(A^2) \frac{A_k A_k}{a^2}, \; k=1,2,3\;\;; \label{diagonal}
\\
T^i_{\;\;0}&=& 0\;\;;
\\
T^i_{\;\;j} &=& \frac{\dot{A_i} \dot{A_j}}{a^2} + 2V'(A^2)\frac{A_i A_j}{a^2}, \;\;\; i\neq j \;\;.
\label{nondiagonal}
\end{eqnarray}
Notice that in the definition of the pressures $p_k$, no summation in $k$ is assumed.

Now we can express the conservation law \mbox{$T^{\mu \nu}_{\;\;\; ;\nu} = 0$} as:
\begin{equation}
\dot{\rho} + H\left( \sum_k p_k + 3 \rho\right) = 0\;. \label{conservation}
\end{equation}
We see that the off-diagonal part of the EMT does not contribute in (\ref{conservation}) because of the homogeneity of the vector field.

The temporal equation (\ref{temporal}), implies\footnote{There is another possible solution: $V'(A^2)=0$.
In this case, the isotropy theorem cannot be applied because the evolution of $A_i$ is not rapid.
It can be showed that $\dot{A}_i=C_i/a$, and the anisotropic components of the EMT decay as $a^{-4}$  \cite{CHMN}.}
\begin{equation}
 A_0 = 0 \label{A0}\;.
\end{equation}

Therefore the homogeneity condition reduces the problem to the evolution of a 3-vector $\vec A(t)$.
We can define the typical time evolution scale of the spatial component $A_i$ from: 
$\omega_i \sim |\dot A_i/A_i|$. This value coincides with the frequency for a oscillatory
movement. In the case of rapid evolution of $A_i$ in relation to the universe expansion $\omega_i \gg H$,
we can neglect time derivatives of the scale factor so that defining $r_i=A_i/a$ in a time range of order
${\omega_i}^{-1}$, we have:
\begin{equation}
\dot{r_i} = \frac{\dot{A_i}}{a} - H r_i \approx \frac{\dot{A_i}}{a}\;.
\end{equation}
Thus, we can ignore the friction term  and we see that Eq. (\ref{motion}) reduces to the evolution equation of a
point particle with position vector $\vec r=\vec A/a$ in the presence of a central potential $V(-r^2)$.
In particular, this implies that we can make use of the standard classical mechanics results. Thus, since
the potential is central, we will have conservation of the corresponding angular momentum
$\vec L=\vec r \times \dot {\vec r}$, which, in turn, implies that the vector $\vec A(t)$ should evolve
in a fixed plane, orthogonal to $\vec L$.

On the other hand, in the case of rapid evolution we can also write:
\begin{equation}
\frac{\dot{A_i} \dot{A_j}}{a^2} \delta^{i j} \approx (\dot{\overrightarrow{r}})^2 = \dot{r^2} + \frac{L^2}{r^2}\;.
\label{radial}
\end{equation}
Thus, the total energy density can be written as:
\begin{eqnarray}
\rho=\frac{1}{2}\frac{\dot{A_i} \dot{A_j}}{a^2} \delta^{i j}+V(A^2)=\frac{1}{2}\dot r^2+\frac{L^2}{2r^2}+V(-r^2)\;, \label{energy1}
\end{eqnarray}
which can be considered as constant within the short time-scale of variation of the
vector field.

Thus, we are left with the reduced radial problem, i.e. the motion of a particle in the radial dimension in the presence
of the effective potential:
\begin{eqnarray}
V_{eff}(r)=\frac{L^2}{2r^2}+V(-r^2)\;,
\end{eqnarray}
where the value of the constant $L$ is set by the initial conditions. The other constant of motion
associated to this problem is the energy density $\rho_0=V_m$.

As commented before, if the temporal evolution of $A_i$ is sufficiently rapid and we concentrate only in
a time interval of order ${\omega_i}^{-1}$, we can ignore the effect of the universe expansion, i.e.
we can consider $a$ as constant. In this case, we can write
the equation of motion for the vector field as $\ddot A_i= 2V'(A^2)A_i$. Let us define:
\begin{eqnarray}
G_{ij}=\frac{\dot A_i A_j}{a^2}, \;\;\; i,j=1,2,3\;;
\end{eqnarray}
 and take its time derivative:
\begin{eqnarray}
\dot G_{ij}=\frac{\ddot A_i A_j}{a^2}+\frac{\dot A_i \dot A_j}{a^2}\;,
\end{eqnarray}
Using the equations of motion, we get:
\begin{eqnarray}
\dot G_{ij}=  2V'(A^2)\frac{A_iA_j}{a^2}+\frac{\dot A_i \dot A_j}{a^2}\;.
\end{eqnarray}
Integrating this expression in a given time interval $[0,T]$:
\begin{eqnarray}\label{interval}
\frac{G_{ij}(T)-G_{ij}(0)}{T}=\left\langle 2V'(A^2)\frac{A_iA_j}{a^2}\right\rangle+\left\langle\frac{\dot A_i \dot A_j}{a^2} \right\rangle,
\end{eqnarray}
with $ i=1,2,3$. If the motion is periodic and $T$ corresponds to the oscillation period, the left hand side (l.h.s.) vanishes.
If the motion is not periodic, but $A_i$ and $\dot A_i$ are bounded, by taking $T$
sufficiently large, but satisfying $H^{-1}\gg T\gg \omega_i^{-1}$ for any $i$, we can also neglect the l.h.s. of the equation.
In those cases, we have:
\begin{eqnarray}
\left\langle\frac{\dot A_i \dot A_j}{a^2} \right\rangle=-\left\langle 2V'(A^2)\frac{A_iA_j}{a^2}\right\rangle, \;\;\;  i,j=1,2,3\;.
\label{virial}
\end{eqnarray}
By using this result in (\ref{nondiagonal}), we straightforwardly get:
\begin{eqnarray}
\left\langle T^i_{\; j} \right\rangle=0,\;\; i\neq j\;;
\end{eqnarray}
i.e. the average EMT is diagonal. Also, using (\ref{virial}) in
(\ref{diagonal}), we obtain
\begin{eqnarray}
\left\langle p_k\right\rangle\equiv-\left\langle T^k_{\; k} \right\rangle&=&\left\langle \frac{1}{2} \frac{\dot{A_i} \dot{A_j}}{a^2} \delta^{i j} \right\rangle -\left\langle V(A^2) \right\rangle,\nonumber\\
& & \;\;\;\;\;\;\;\;\;\;
\;\; k=1,2,3\;; %\nonumber \\
\end{eqnarray}
i.e. all the average pressures are the same, so that the isotropy of the mean value of the EMT is proved:
\begin{eqnarray}
\left\langle T^i_{\; j} \right\rangle=-\left\langle p \right\rangle\,\delta^{i}_{\;j}\;.
\end{eqnarray}

\begin{figure}[t]
\includegraphics[width=75mm,height=75mm,angle=0]{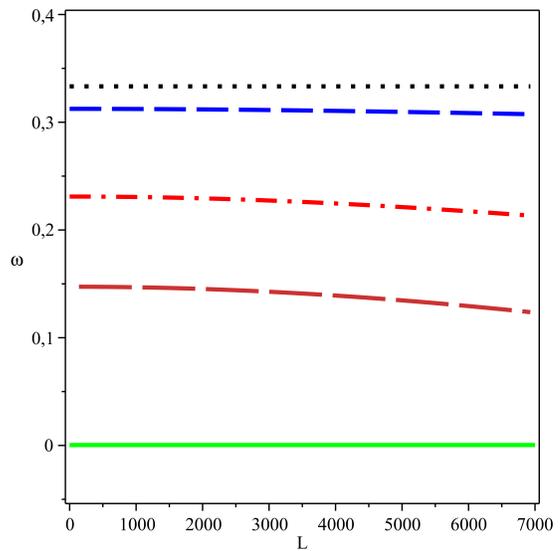}
\caption{The effective equation of state $\omega\equiv \left\langle p \right\rangle/\left\langle \rho  \right\rangle$,
can be computed numerically by integrating expressions (\ref{energy}) and (\ref{diagonal}) with respect to the
$r$ variable with the use of Eq. (\ref{radial}) and (\ref{energy1}). Thus, in the case $V=-a A^2$,
$\omega=0$ and the coherent vector oscillations behave as  cold dark matter (CDM); or for $V=b A^4$, $\omega=1/3$
and they behave as radiation. These results agree with Eq. (\ref{omega}). For a general potential, the
effective equation of state depends on the initial conditions: $V_m$ and $L$. In this figure, one can see the
$L$ dependence of $\omega$ for $V=-a A^2+b A^4$ and $V_m=10^4$. Depending on the particular values of $a$
and $b$, the equation of state interpolates between the radiation and the CDM behavior: from top to bottom respectively,
dotted (black) line corresponds to $(a,b)=(10^{-7},2.4 \cdot 10^{-7})$, blue (dashed)
line corresponds to $(a,b)=(0.1,2\cdot 10^{-4})$, dashed-dotted (red) for
$(a,b)=(0.4, 10^{-4})$, long dashed (orange) for  $(a,b)=(0.7, 6\cdot 10^{-5})$ and
continuous (green) for  $(a,b)=(10^{-2},10^{-11})$.}
\end{figure}

It is interesting to note that the virial relations (\ref{virial}) also allow us to get the average equation
of state for a power-law potential: $V=\lambda (A_\mu A^\mu)^n$. In this case, we have:
\begin{eqnarray}\label{viri}
\left\langle\frac{1}{2}\frac{\dot A_i \dot A_j}{a^2}\delta^{ij} \right\rangle= n\left\langle V(A^2)\right\rangle\;,
\end{eqnarray}
what implies:
\begin{eqnarray}
\left\langle p_k\right\rangle=\left\langle p\right\rangle = (n-1)\left\langle V(A^2) \right\rangle,
\;\; k=1,2,3\;.
\end{eqnarray}
On the other hand, for the average energy-density, we get from (\ref{energy}):
\begin{eqnarray}\label{rhov}
\left\langle \rho\right\rangle= (n+1)\left\langle V(A^2) \right\rangle\;,
\end{eqnarray}
and finally:
\begin{eqnarray}\label{omega}
\omega=\frac{\left\langle p\right\rangle}{\left\langle \rho\right\rangle}=\frac{n-1}{n+1}\;,
\end{eqnarray}
which agrees with the scalar case \cite{Turner} and shows that the average equation of state does
not depend on the polarization of the vector oscillations (notice that the scalar
case corresponds to $L=0$, i.e the evolution of a particle in one spatial dimension).  In the case of a general potential,
not necessarily of the power-law form, the equation of state cannot be obtained analytically and,
in general, it will depend on the initial conditions for the vector oscillations, as Fig. 1 shows.
Notice that the average energy density and pressure satisfy the conservation equation
(\ref{conservation}) up to corrections of order $\Od(HT)$.

We could think  that the isotropy of the average EMT could have been inherited
from the isotropy of the FLRW metric. But repeating the process for a Bianchi I metric:
%$ds^2 = dt^2 - a^2_1 ( t )  dx^2 -a^2_2(t) dy^2- a^2_3 ( t ) dz^2$,
\begin{eqnarray}
ds^2 = dt^2 - a^2_1 ( t )  dx^2 -a^2_2(t) dy^2- a^2_3 ( t ) dz^2\;,
\end{eqnarray}
if we  replace
\begin{eqnarray}
\frac {A_i}{a} \rightarrow \frac{A_i}{a_i} , \; i=1,2,3\;;
\end{eqnarray}
and assume $\omega_i$ much greater than $H_i=\dot a_i/a_i$, it is straightforward to obtain the
same results for $\langle T^{\mu}_{\;\;\nu}\rangle$.

The previous results can be directly extended
to  general space-time geometries (not necessarily homogeneous). With that
purpose, let us consider a locally inertial observer at
$x_0^\mu=0$ and write
the metric tensor in Riemann normal coordinates around $x_0^\mu$ \cite{Petrov}:
\begin{eqnarray}
g_{\mu\nu}(x)=\eta_{\mu\nu}+\frac{1}{3}R_{\mu\alpha\nu\beta}x^\alpha x^\beta +\dots
\label{normal}
\end{eqnarray}
Let assume that the following conditions hold:
\begin{enumerate}
\item {The Lagrangian of the vector field is restricted to be given by the form of Eq. (\ref{L}).}

\item {The vector field evolves rapidly:
\begin{eqnarray}
|R^\gamma_{\lambda\mu\nu}| \ll \omega_i^2 ,\;
\text{and} \;\;
|\partial_j A_i| \ll | \dot{A}_i | ,\;\; i,j=1,2,3
%\nonumber \\
%&&\;\;\;\;\;\;\;\;\text{for}\;\; i,j=1,2,3\;,
\end{eqnarray}
for  any component of the Riemann tensor.}

\item {$A_i$ and $\dot A_i$ remain bounded in the evolution.}
\end{enumerate}
The second condition implies that if we are only interested in time scales of
order $\omega_i^{-1}$, then we are in a normal neighborhood and we can neglect the second
term in (\ref{normal}) and also work with a homogeneous vector field.   In such a case,  it is possible to
rewrite all the above equations with $a=1$. Thus, by using an interval $[0,T]$ that verifies the condition:
\begin{eqnarray}
|R^\gamma_{\lambda\mu\nu}| \ll T^{-2} \ll \omega_i^2\;,
\end{eqnarray}
for any of the components of the Riemann tensor and any of the spatial components of the vector: $i=1,2$ and $3$,
it is possible to obtain (\ref{interval}) and prove that the mean value of the EMT is isotropic. It is
interesting to note that if the metric is non-homogeneous, the EMT can be non-homogeneous, but its
average value will be isotropic as seen from a locally inertial frame.

In this work, we have considered the evolution of general Abelian vector fields in an expanding universe.
We have shown by means of a mechanical analogy with a particle moving in a central potential that
a generalized version of the  virial theorem ensures that, for rapid dynamics, the average EMT is always diagonal and isotropic,
despite the fact that the evolution always takes place in a fixed plane. The result
 can be extended to arbitrary geometries within a normal neighborhood of any
locally inertial observer.

%\vspace{0.2cm}

{\bf Acknowledgements}
This work has been supported by MICINN (Spain) project numbers FIS 2008-01323, FIS2011-23000, FPA2011-27853-01 and Consolider-Ingenio MULTIDARK CSD2009-00064.

\end{document}